# Identifying Crosscutting Concerns Using Fan-in Analysis

Marius Marin, Leon Moonen and Arie van Deursen

**TU**Delft

SERG

# Identifying Crosscutting Concerns Using Fan-in Analysis

MARIUS MARIN

Delft University of Technology

ARIE VAN DEURSEN

Delft University of Technology & CWI

and

LEON MOONEN

Delft University of Technology

Aspect mining is a reverse engineering process that aims at finding crosscutting concerns in existing systems. This paper proposes an aspect mining approach based on determining methods that are called from many different places, and hence have a high *fan-in*, which can be seen as a symptom of crosscutting functionality. The approach is semi-automatic, and consists of three steps: metric calculation, method filtering, and call site analysis. Carrying out these steps is an interactive process supported by an Eclipse plug-in called FINT. Fan-in analysis has been applied to three open source Java systems, totaling around 200,000 lines of code. The most interesting concerns identified are discussed in detail, which includes several concerns not previously discussed in the aspect-oriented literature. The results show that a significant number of crosscutting concerns can be recognized using fan-in analysis, and each of the three steps can be supported by tools.

## 1. INTRODUCTION

Aspect-oriented software development (AOSD) is a programming paradigm that addresses *crosscutting concerns*: features of a software system that are hard to isolate, and whose implementation is spread across many different modules. Well-known examples include logging, persistence, and error handling. Aspect-oriented programming captures such crosscutting behavior in a new modularization unit, the *aspect*, and offers code generation facilities to *weave* aspect code into the rest of the system at the appropriate places.

*Aspect mining* is an upcoming research direction aimed at finding crosscutting concerns in existing, non-aspect oriented code. Once these concerns have been identified, they can be used for program understanding or refactoring purposes, for example by integrating

Author's addresses: Marius Marin, Software Evolution Research Lab, EEMCS, Delft University of Technology, The Netherlands, A.M.Marin@tudelft.nl. Arie van Deursen, Software Evolution Research Lab, EEMCS, Delft University of Technology & CWI, The Netherlands, Arie.vanDeursen@tudelft.nl. Leon Moonen, Software Evolution Research Lab, EEMCS, Delft University of Technology, The Netherlands, Leon.Moonen@computer.org.
This is a substantially revised and extended version of [Marin et al. 2004a].

aspect mining techniques into the software development tool suite. In addition to that, aspect mining research increases our understanding of crosscutting concerns: it forces us to think about under what circumstances a concern should be implemented as an aspect, it helps us find crosscutting concerns that are beyond the canonical ones such as logging and error handling, and it may lead to concerns that are crosscutting, yet not easily modularized with current aspect technology (such as, e.g., ASPECTJ).

In this paper we propose *fan-in analysis*, an aspect mining approach that involves looking for methods that are called from many different call sites and whose functionality is needed across different methods, potentially spread over many classes and packages. Our approach aims at finding such methods by computing the fan-in metric for each method using the system's static call graph. It relies on the observation that scattered, crosscutting functionality is likely to generate high fan-in values for key methods implementing this functionality. Furthermore, it is consistent with the guidelines of applying aspect solutions when the same functionality is required in many places throughout the code [Colyer et al. 2005].

Fan-in analysis is a semi-automated process consisting of three steps. First, we identify the methods with the highest fan-in values. Second, we filter out methods that may have a high fan-in but for which it is unlikely that there is a systematic pattern in their usage that could be exploited in an aspect solution. Typical examples are getters and setters, as well as utility methods. Third, we inspect the call sites of the high fan-in methods, in order to determine if the method in question does indeed implement crosscutting functionality. This step is the most labor intensive, and it is based on an analysis of recurring patterns in, for example, the call sites of the high fan-in method. All steps are supported by an Eclipse[1] plug-in called FINT, which is also discussed in the paper.

We discuss the application of fan-in analysis to three existing open source systems (the web shop PETSTORE, the drawing application JHOTDRAW, and the servlet container TOMCAT) implemented in Java. For all systems our approach found a number of interesting crosscutting concerns that could benefit from an aspect-oriented redesign.

When evaluating the quality of an aspect mining technique, two challenges have to be faced. The first is that a benchmark system must exist in which the crosscutting concerns are known already, for example because they have been identified by an expert. At the moment, such a benchmark does not exist. A growing number of aspect mining researchers, however, are using JHOTDRAW as their case study, which is thus evolving into such a benchmark system.

The second evaluation challenge is that the decision that a concern is crosscutting and amenable to an aspect-oriented implementation is a design choice, which is a trade-off between alternatives. Thus, there is not a yes/no answer to the question whether a concern identified is suitable for an aspect implementation. As a consequence, quantitative data on the number of false negatives (how many crosscutting concerns are missed) or false positives (how many of the concerns we identified are in fact not crosscutting) has a subjective element to it. This means that an evaluation of an aspect mining technique just in terms of, for example percentages of false positives and negatives, or in terms of precision and recall, is an oversimplification.

To deal with these issues, we decided to discuss a substantial number of concerns found in considerable detail, explaining for what reasons they should be considered as cross-cutting concerns. In order to encourage a debate on our results, we selected open source systems on purpose, allowing others to see all code details when desired.

---

[1] www.eclipse.org

As a result, the paper can be read in two ways. First of all, it is the presentation of the fan-in aspect mining technique. Second, it is a discussion of those crosscutting concerns that were found in three open source systems by means of fan-in analysis – thus establishing a first step towards a common benchmark that can be used in further aspect mining research.

The scope of the present paper is aspect mining itself. Using the aspect mining results, for example for refactoring to ASPECTJ, is a separate topic, for which we refer to, e.g., Binkley et al. [2005], as well as to our own work on reimplementing some of the concerns discussed in the present paper [van Deursen et al. 2005; Marin et al. 2005].

This paper is organized as follows. We start out by surveying existing work in the area of aspect mining. Then, in Section 3, we present the fan-in metric, the analysis steps, as well as the Eclipse plug-in supporting fan-in analysis. In Section 4 we present an overview of the case studies. In Sections 5–7 we cover the results obtained from applying fan-in analysis to three open source case-studies presenting several of the concerns found in considerable detail. We reflect on these case studies, on the reasons for success, and on the limitations of our approach in Section 8. We conclude with a summary of the paper's key contributions and opportunities for future work.

We assume the reader has basic knowledge of aspect-oriented programming, and we refer to Gradecki and Lesiecki [2003], The AspectJ Team [2003], and Laddad [2003b] for more information.

## 2. ASPECT MINING: BACKGROUND AND RELATED WORK

Since aspect mining is a relatively recent research area, we start out by providing some uniform terminology. We then discuss the most important aspect mining approaches published to date.

### 2.1 Terminology

Sutton and Rouvellou [2005] provide a discussion on what constitutes a "concern". Following them, we take concern generally to be "any matter of interest in a software system." Concerns can live at any level, ranging from requirements, to use cases, to patterns and contracts. In this paper we will focus on concerns that play a role at the source code level.

We distinguish between a concern's *intent* and *extent*:

—A concern's *intent* is defined as the objective of the concern. For example, the intent of a tracing concern is that all relevant input and output parameters of public methods are appropriately traced.

—A concern's *extent* is the concrete representation of that concern in the system's source code. For example, the extent of the tracing concern consists of the collection of all statements actually generating traces for a given method parameter.

In aspect mining, we search for source code elements that belong to the extent of concerns that *crosscut* the software system's modularization structure. Such *crosscutting concerns* are not dedicated to a modularization unit like a single package, class hierarchy, class, method, but are *scattered* over all these units. As an example, the tracing concern will affect many different methods distributed over different packages or classes. A consequence of this scattering is *tangling*: modular units cannot deal exclusively with their core

```
package myaspects;
public aspect Tracing {

    declare parents: mypackage.* implements Traceable ;

    public interface Traceable {
        public void traceEntry(String methodSig);
        public void traceExit(String methodSig);
    }

    public void Traceable.traceEntry(String methodSig) {
        System.out.println("Entering " + methodSig);
    }

    public void Traceable.traceExit(String methodSig) {
        System.out.println("Exiting " + methodSig);
    }

    pointcut thePublicMethods(Traceable t) :
        target(t) &&
        execution(public * mypackage..*(..)) &&
        !within(Tracing );

    Object around(Traceable t): thePublicMethods(t) {
        t.traceEntry(thisJoinPoint.getSignature().toString());
        Object result = proceed(t);
        t.traceExit(thisJoinPoint.getSignature().toString());
        return result;
    }
}
```

Fig. 1.    ASPECTJ definition for the tracing concern

concern, but have to take into account the implementation of other concerns that crosscut their modularization as well.

Aspect-oriented software development aims at avoiding the maintenance problems caused by scattering and tangling by making use of the new aspect modularization construct. As a simple example, consider an implementation of the tracing concern in AS-PECTJ[2], as shown in Figure 1. The *declare* statement at the top of the aspect body ensures that all classes contained in a particular package extend the *Traceable* interface, using a so-called inter-type declaration. The *Traceable* interface itself is provided in the subsequent lines, including a default implementation of the interface. In this way, the aspect extends multiple classes, thereby capturing the statically crosscutting nature of tracing. The remainder of the aspect captures the dynamic crosscutting, using a "pointcut" which intercepts all calls to public methods, and "around advice" that emits a string with the signature of the executing method just before and just after its execution. The aspect can be woven into the base code, keeping the latter *oblivious* to the tracing concern. This helps to reduce the tangling in the base code and provides a non-scattered implementation of the crosscutting concern. Furthermore, a (small) reduction in code size can be achieved if the crosscutting is sufficiently regular (as is the case with the tracing concern: the pointcut expression can quantify over all public methods).

---

[2] www.aspectj.org

Aspect mining aims at finding crosscutting concerns in existing, non-aspect-oriented code. Such concerns could possibly be improved by applying aspect-oriented solutions or can be documented for program comprehension purposes. The mining involves the search for source code elements belonging to the implementation of a crosscutting concern, i.e., which are part of the concern's extent. We will refer to such code elements as *seeds*. Once we have found a single seed for a concern, we can try to expand the seed to the full extent of the concern, for example by following data or control flow dependencies.

Aspect mining generally requires human involvement. Therefore, we will say that aspect mining tools yield *candidate seeds*, which can be turned into *confirmed seeds* (or simply "seeds") if accepted by a human expert, or *non-seeds* if rejected. Sometimes a non-seed is also referred to as a *false positive* – a *false negative* then is a part of a known crosscutting concern, potentially detectable by the technique, but missed due to inherent limitations of the approach or due to the specific filters applied in it. The key aspect mining challenge is to keep the percentage of confirmed seeds in the total set of candidate seeds as high as possible, without increasing the number of false negatives too much. As we will see, with fan-in analysis this percentage is above 50%.

The origins of aspect mining can be traced back to the concept assignment problem, i.e., the problem of discovering domain concepts and assigning them to their realizations within a specific program [Biggerstaff et al. 1994]. Work on this problem has resulted in such research areas as feature location [Koschke and Quante 2005; Wilde and Scully 1995; Xie et al. 2006], design pattern mining [Ferenc et al. 2005], and program plan recognition [Rich and Wills 1990; Wills 1990; van Deursen et al. 2000].

In aspect mining we specifically search for concerns (concepts) whose realization in a given program cuts across modular units. Several aspect mining approaches have been published, for which we propose a distinction between *query-based* and *generative* approaches. *Query-based* approaches start from manual input such as a textual pattern. *Generative* approaches, including fan-in analysis, aim at generating seeds automatically making use of, for example, structural information obtained from the source code. Below we discuss these two categories of aspect mining approaches. Moreover, we discuss techniques that are most closely related to our fan-in analysis.

### 2.2   Query-Based Approaches

Query-based, explorative techniques rely on search patterns provided by the user. Source code locations that match the pattern correspond to crosscutting concern seeds, which can subsequently be expanded to more complete concerns using a tool.

One of the first query-based tools, the Aspect Browser, uses lexical pattern matching for querying the code, and a map metaphor for visualizing the results [Griswold et al. 2001]. The Aspect Mining Tool AMT extends the lexical search from the Aspect Browser with structural search for usage of types within a given piece of code [Hannemann and Kiczales 2001]. Both tools display the query results in a Seesoft-type view as highlighted strips in enclosed regions representing modules (e.g., compilation units) of the system [Eick et al. 1992].

AMTEX is an AMT extension that provides support for quantifying the characterization of particular aspects [Zhang and Jacobsen 2003]. AMTEX, in turn, has evolved into PRISM, a tool supporting identification activities by means of lexical and type-based patterns called *fingerprints* [Zhang and Jacobsen 2004]. A fingerprint can be defined, for example, as any method in a given class of which the name starts with a given word. A software engineer defining fingerprints is assisted by so-called *advisors*. PRISM currently provides a ranking advisor which reports the most frequently-used types across methods. This idea is akin to fan-in analysis, which reports the most frequently used methods across a system. There are, however, no reports about the successfulness of applying the approach implemented in PRISM to the identification of crosscutting concerns.

The Feature Exploration and Analysis Tool FEAT is an Eclipse plug-in aimed at locating, describing, and analyzing concerns in source code [Robillard and Murphy 2002]. It is based on *concern graphs* which represent the elements of a concern and their relationships. A FEAT session starts with an element known to be a concern seed, and FEAT allows the user to query relations, such as direct call relations, between the seed and other elements in the program. The results of the query that are considered relevant by the user to the implementation of a (crosscutting) concern can be added to the graph-based representation of the concern.

The Concern Manipulation Environment CME aims at providing support across the whole lifecycle of an aspect-oriented development project [Harrison et al. 2004]. This support also includes aspect identification facilities through an integrated search component (Puma) that uses an extensible query language (Panther) [Tarr et al. 2004]. The Panther language includes the static part of the AspectJ pointcut language. CME also allows for concern management similar to FEAT. Most importantly, CME provides a possible infrastructure for the integration of different approaches to aspect mining, including seed identification and concern exploration and management.

Various query-based tools (the Aspect Browser, AMT, and FEAT) have been compared in a recent study [Murphy et al. 2005]. This study shows that the queries and patterns are mostly derived from application knowledge, code reading, words from task descriptions, or names of files. As the study shows, prior knowledge of the system or known starting points strongly affect the usefulness of the outcomes of the analysis.

## 2.3 Generative Approaches

The second group of aspect mining approaches aim at automatically generating crosscutting concern seeds with a good quality: seeds that will reduce the effort of further understanding and exploring the concern. The approaches in this category can be described as *generative* techniques and will typically provide the input for the explorative approaches.

Many generative approaches use program analysis techniques to look for symptoms of code scattering and tangling and identify code elements exhibiting these symptoms that can act as candidate aspect seeds.

Shepherd et al. [2004] use clone detection based on program dependence graphs and the comparison of individual statement's abstract syntax trees for mining aspects in Java source code.

Three clone detection tools, implementing matching on tokens, abstract syntax trees, and on program dependence graphs, respectively, are evaluated by Bruntink et al. [2005] on an industrial C component. The starting point were four dedicated crosscutting concerns that were manually identified and annotated in the code beforehand. The evaluation assesses the suitability of clone detection for identifying these concerns automatically by measuring the coverage of the annotated concerns by detected clones.

Code clones in object-oriented systems would typically be refactored through method extraction [Fowler et al. 1999] which results in scattered calls to the extracted method [Laddad 2003a]. Fan-in analysis looks for the concerns implemented by these scattered calls, which could be further refactored into aspect advice.

Dynamic analysis has been considered for aspect identification by examining execution traces for recurring execution patterns [Breu and Krinke 2004] and by applying formal concept analysis to associate method executions to traces specific to documentation-derived use-case scenarios [Tonella and Ceccato 2004a]. Particularly challenging for dynamic analysis techniques is to exercise all functionality in the system that could lead to aspect candidates. This implies that a preliminary activity is needed in which use-case scenarios are defined for the system under investigation. Fan-in analysis does not require such a preliminary activity.

The first of the two dynamic techniques has been adapted recently to static analysis to search for recurring execution patterns in control flow graphs [Krinke 2006]. The technique is similar in some respect to fan-in analysis, which searches for recurrent call relations. The experimental results of the technique are discussed by comparison with our own results reported for one of the analyzed systems, and show many common findings.

Formal concept analysis has also been applied in an identifier analysis that groups programming elements based on their names [Tourwé and Mens 2004]. This analysis starts from the assumption that naming conventions can be used to relate the scattered elements of a concern. Although fan-in analysis could use naming conventions for the investigation of the automatically generated results, its primary functionality relies on structural relationships.

The suitability of refactoring certain interfaces implemented by a class has been investigated through a number of indicators like the naming pattern used by the interface definition, the coupling between the methods of the implementing class and the methods declared by the interface, or the package location of the interface and its implementing class [Tonella and Ceccato 2004b]. By comparison with fan-in analysis which focuses on method seeds, this technique is directly targeting interface definitions for seed identification.

Besides our own experiments [Marin et al. 2004a], assessments of fan-in analysis have been provided by Gybels and Kellens [2005] who used the metric as an approximate heuristic for measuring scattering. Another assessment of this analysis has been made available through the Timna framework [Shepherd et al. 2005] which uses machine learning techniques to combine the results of several aspect mining techniques.

In their more recent work, Breu and Zimmermann [2006] search for concerns by analyzing the changes in the values of the fan-in metric between different versions of the system under investigation. The technique they propose examines the version history for insertions of method calls. Similar to fan-in analysis, a reported seed consists of a set of one or more methods with same call site locations. This technique could complement fan-in analysis by giving insight into the evolution of the metric's values in a system, and hence into the evolution of the concern of a method.

### 2.4   Aspect Identification Case Studies

The subject systems that we have analyzed in the previous [Marin et al. 2004a] and present work have also been used by related research [Shepherd et al. 2005; Shepherd et al. 2004; Janzen and De Volder 2003; Binkley et al. 2005] or in tool demonstrations (e.g., FEAT [Robillard and Murphy 2002]). However, our work on fan-in analysis is the first attempt to establish a common benchmark for the development of aspect mining techniques, by explicitly reporting the results obtained for a number of case-studies and discussing them in significant detail. This work has been continued in a comparative study [Ceccato et al. 2006] of the fan-in technique with the dynamic [Tonella and Ceccato 2004a] and identifier analysis [Tourwé and Mens 2004] approaches. The JHotDraw case-study targeted by the comparison experiment is intended to become the de-facto benchmark for aspect mining.

### 3.   ASPECT MINING USING FAN-IN ANALYSIS

Fan-in analysis fits in the category of generative aspect mining approaches. The main symptom of crosscuttingness it tries to capture is *scattering*: the code for one concern is spread across the system. If the scattered pieces of code have functionality in common, it is likely that this will have been factored out in helper methods. These methods are then called from many places, giving them a high fan-in value. In an aspect-oriented re-implementation of such concerns, the method would constitute (part of) the advice, and the call site would correspond to the context that needs to be captured using a pointcut.

Fan-in analysis consists of three steps:

(1)  Computation of the fan-in metric for all methods;

(2)  Filtering of the set of methods to obtain the methods that are most likely to implement crosscutting behavior;

(3)  Analysis of the remaining methods to determine which of them are part of the implementation of a crosscutting concern.

The next subsections describe each of these steps, as well as the tool FINT supporting these steps.

### 3.1   A Fan-In Metric for Aspect Mining

The metric we will use for aspect mining is based on method fan-in, which is a "measure of the number of methods that call some other method" [Sommerville 2004]. Thus, we will collect the set of (potential) callers for each method — and the cardinality of this set gives the required fan-in value. The actual value, however, of method fan-in depends on the way we take polymorphic methods (callers as well as callees) into account.

Therefore, our first refinement is that we count the number of *different method bodies* that call some other method. Thus, if a single abstract method is implemented in two concrete subclasses, we treat these two implementations as separate callers.

Our second refinement deals with calls to polymorphic methods. Recall that we are interested in methods that are called from many different places, since these are potentially part of a crosscutting concern. If we find that a particular method $m$ belongs to such a concern, it is very likely that superclass declarations or subclass overrides of $m$ belong to that same concern. For that reason, if we see that method $m'$ applies method $m$ to an object of static type $C$, we add $m'$ to the set of (potential) callers for each $m$ declared in any sub- or superclass of $C$.

With this definition, (abstract) method declarations high in the inheritance hierarchy act as fan-in accumulators: whenever a specific subclass implementation is explicitly invoked, the fan-in of not only the specific but also of the abstract method is increased. In this way, if there are many calls to different specific implementations, we get a high fan-in value for the superclass method. An aspect-oriented reimplementation would aim at capturing the many specific call sites into a pointcut, and invoke the abstract method in the advice, relying on polymorphism to dispatch to the proper specific implementation.

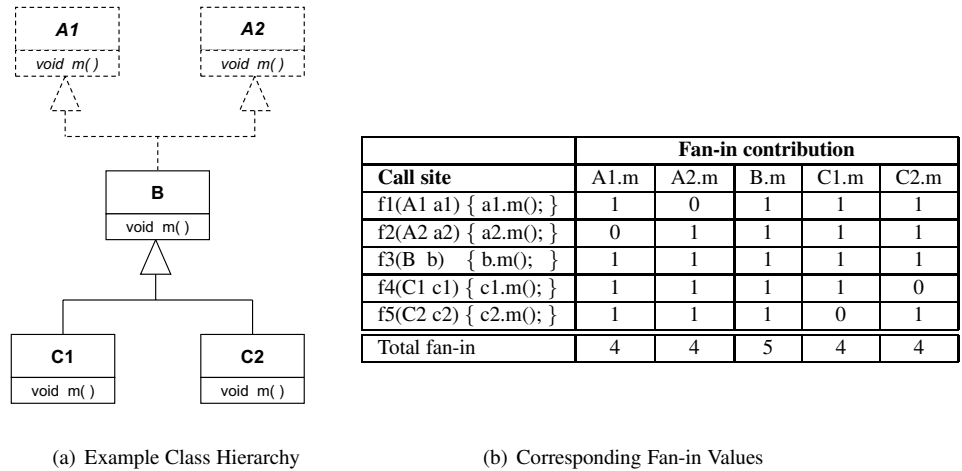

(a) Example Class Hierarchy

(b) Corresponding Fan-in Values

|  | Fan-in contribution | | | | |
|---|---|---|---|---|---|
| Call site | A1.m | A2.m | B.m | C1.m | C2.m |
| f1(A1 a1) { a1.m(); } | 1 | 0 | 1 | 1 | 1 |
| f2(A2 a2) { a2.m(); } | 0 | 1 | 1 | 1 | 1 |
| f3(B  b)   { b.m(); } | 1 | 1 | 1 | 1 | 1 |
| f4(C1 c1) { c1.m(); } | 1 | 1 | 1 | 1 | 0 |
| f5(C2 c2) { c2.m(); } | 1 | 1 | 1 | 0 | 1 |
| Total fan-in | 4 | 4 | 5 | 4 | 4 |

Fig. 2.   Example class hierarchy and corresponding fan-in values

An example hierarchy is shown in Figure 2. The example illustrates the effects of various calls to a polymorphic method *m* in different positions in the class hierarchy. Note that, given our definition, the fan-in for method m in class C1 is not affected by calls to m defined in C2 and vice versa: the same holds for sibling classes A1 and A2.

Our last refinement is concerned with super calls. For super calls, we explicitly know which method is targeted, which therefore is the only method whose call set is extended.

Observe that there are multiple ways in which a fan-in metric can be defined. Historically, the notion of fan-in was introduced by Henry and Kafura [1981] as an indicator for coupling in procedural software. They include data access in fan-in as well, which we do not. An overview of coupling indicators for object-oriented systems is discussed by Briand et al. [1999]. In some cases these metrics are based on a derivative of the fan-in metric, which then often is taken at the class level (instead of the method fan-in we use) – see, e.g., Henderson-Sellers et al. [1996]. In other cases calls from private methods are excluded from the fan-in count.

### 3.2   Method Filtering

After having computed the fan-in values of all methods, we apply the following filters, in order to obtain a smaller set of methods with a higher chance of implementing crosscutting behavior.

First, we restrict the set of methods to those having a fan-in above a certain threshold. This can be an absolute fan-in value (say, 10) or a relative percentage (say, the top 5% of all methods ordered by their fan-in values). Note that an absolute value threshold not only acts as a filter, but also an indicator for the severity of the scattering.

In our case studies, we experimented with several values, and found 10 to be a useful trade-off between the number of concerns that one can find and the number of methods that need to be inspected.

Second, we filter getters and setters from the list of methods. This is either based on naming conventions (methods matching the "get*" or "set*" pattern) or on an analysis of the method's implementation.

Last but not least, we filter utility methods, like `toString()`, classes such as *XMLDocumentUtils* containing "util" in their name, collection manipulation methods, and so on, from the remaining set. This is a manual step that may require some familiarity with the subject system. This familiarity can be improved after each iteration by looking at the results and analyzing apparent indicators like names or easily accessible documentation, such as descriptive comments in the code. The heuristics we used for identifying utility methods in our case studies are based on the following categories:

—Methods that belong to collection classes and/or packages. The JHOTDRAW case study, for example, comes with its own library for collection classes. We typically recognized these based on class or package names, such as *FigureEnumerator*, *HandleEnumerator*, *ListWrapper*, and so on.

—Documented utilities, based on naming and easily available documentation criteria. For example, for PETSTORE, the utility methods belong to two classes: *XMLDocumentUtils* and *PopulateUtils*, which creates and prints SQL statements used to populate the sample database for the application. In TOMCAT, we marked classes from the *util.buf* package as utility, which deals with encoding and decoding buffers. We also marked the *util.digester.Digester* class as utility - the class is described as an XML parser in Tomcat's documentation.

### 3.3 Seed Analysis

Our final step is to conduct a manual analysis of the remaining set of methods. This analysis follows a number of guidelines, part of which benefit from automatic support. Reasoning about the reported candidates can take a top-down or bottom-up approach.

In the bottom-up approach we look for consistent invocations of the method with a high fan-in value from call sites that could be captured by a pointcut definition. Examples of such consistent invocations include:

—The calls always occur at the beginning or the end of a method;

—The calls occur in methods that are all refinements of a single abstract method, as, for instance, for contracts exercised across class hierarchies;

—The calls occur in methods with similar names, like handlers for mouse or key events;

—All calls occur in methods implementing a certain role, as, for example, listener-objects that register themselves as observers of a subject-object state.

The regularity of these call sites typically will make it possible to capture the calls in a pointcut mechanism, and the high fan-in method into advice. The main challenge of the bottom-up approach is to recognize these patterns leading to pointcuts. As we will see in the next section, it is possible to offer tool support here that helps the human engineer in conducting this analysis.

In the top-down approach, we take domain knowledge or knowledge of typical crosscutting concerns into account, as described by, e.g., Hannemann and Kiczales [2002] or Laddad [2003b]. For example, a number of design patterns define (crosscutting) roles and methods specific to these roles that can appear in the list of seeds. The human engineer can take advantage of such knowledge when analyzing the candidate seeds to recognize the pattern-specific roles. The Composite pattern, for example, defines roles and methods to allow parent-objects to refer and manipulate child-elements. Similarly, the methods in a decorator class are characterized by the consistent redirection functionality they implement.

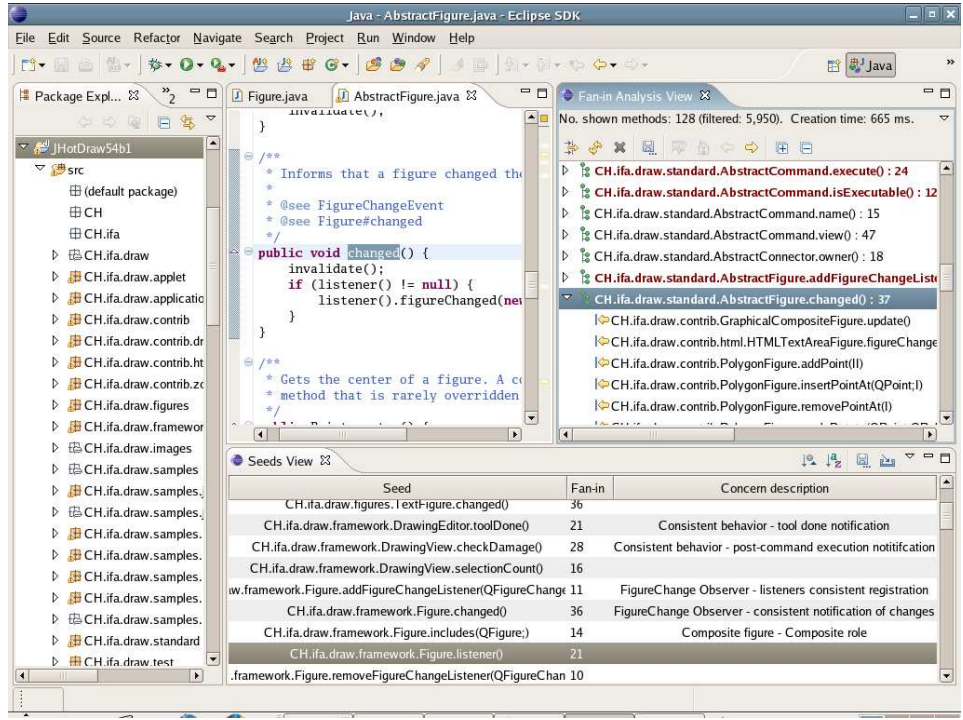

Fig. 3.    FINT in action, showing the *Fan-in Analysis View* (top right) and the *Seeds View* (bottom right).

### 3.4    The Fan-in Tool FINT

The Fan-in Tool FINT[3] is an Eclipse plug-in that provides automatic support for the metric computation, method filtering, and candidate analysis steps of fan-in analysis.

To compute the fan-in metric, the tool first builds the abstract syntax tree for the user-selected sources, and then creates a call graph with the methods declared in the selected sources and their callees. The fan-in metric is derived from this graph, as described in Section 3.1. The results are displayed in the *Fan-in Analysis view*, shown in Figure 3, together with the list of callers for each method. The results can be ordered alphabetically or by their fan-in value. Optionally, the results can also be stored on file.

The same view is used for the filtering step of fan-in analysis. Thus, the user can indicate an absolute threshold for the fan-in value. Furthermore, the user can choose to filter out accessor methods by their signature based on the "get*" or "set*" naming convention, or based on their implementation.

---

[3] `http://swerl.tudelft.nl/bin/view/AMR/FINT`. The features discussed in this paper are part of FINT 0.6.

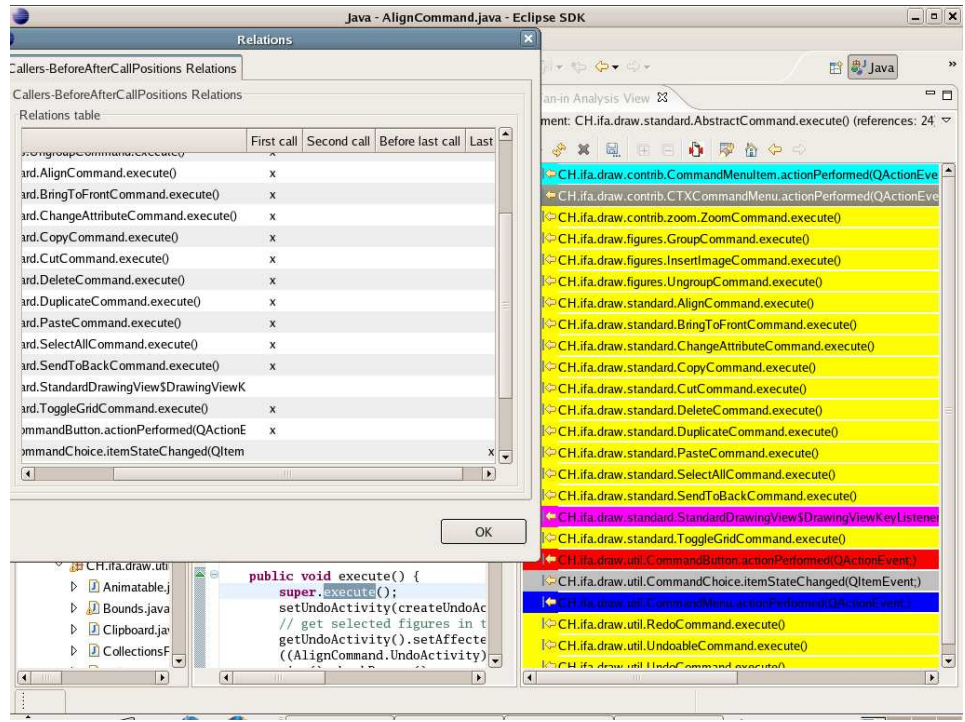

Fig. 4. Seed inspection using FINT. The color codings in the right window indicate inheritance from common interfaces; the table at the left marks the positions of calls to a high fan-in method.

In addition to that, the user can indicate groups of elements whose methods are to be excluded from the callee or caller sets. Excluded callees are indicated as utility-methods and represent methods considered irrelevant for analysis. Similarly, the user-selected callers will not contribute to the fan-in metric of the analyzed methods. Both filters can be applied, for instance, to (JUnit) tests, which are neither relevant as candidate-seeds nor as callers. The user marks these elements in a browser window, which displays the Java elements in the hierarchy of the analyzed elements, similar to Eclipse's *Package Explorer* view. The user can select a check-box for the enclosing package, file, or declaring class of the method to be filtered.

Methods not declared in the analyzed sources, but called by analyzed methods are considered *libraries* and can optionally be included in the analysis. These methods cannot contribute to the fan-in metric of a method.

The *Fan-in Analysis View* is also the starting point for the last analysis step. From this view, the engineer can inspect the reported methods and their callers. Methods can be marked as seeds and added to the *Seeds View*, shown at the bottom of Figure 3. In this view, the seeds can be documented with a concern description, saved to a file or loaded from a previous analysis.

The analysis and seed views from FINT support the user in recognizing recurring patterns and similarities as discussed in the previous section, helping him or her in deciding whether one or more high fan-in methods belong to a crosscutting concern.

|  | PETSTORE | JHOTDRAW | TOMCAT |
|---|---|---|---|
| size in non-comment lines of code | 17,032 | 20,594 | 149,219 |
| number of methods | 1,917 | 3,230 | 13,489 |
| methods with fan-in $\geq 10$ | 16 (1%) | 205 (6%) | 424 (3%) |
| | | | |
| Statistics for methods with fan-in $\geq 10$ | | | |
| utility methods | 3 | 20 | 16 |
| accessors | 5 | 71 | 181 |
| confirmed seeds | 7 (87%) | 58 (51%) | 164 (73%) |
| non-seeds | 1 (13%) | 56 (49%) | 63 (27%) |
| concerns | 5 | 10 | 10 |

Table I.   Key statistics of our case studies

The various ways in which methods and call sites can be sorted and inspected in FINT help to discover such patterns. Furthermore, the tool provides automatic support for detecting some of the possible relations between the callers of an analyzed method, like grouping of the callers by common hierarchies or their declaring interfaces, by the position of the analyzed call, or by other callees shared by the callers.

As an example, Figure 4 shows the view for analyzing the callers of a method with a high fan-in value by investigating their declaring interfaces. The callers declared by the same interface are shown in a same, distinctive color. Such analysis is helpful, for example, in identification of (crosscutting) responsibilities that are to be fulfilled by a number of classes.

The same figure also shows a relational table for the callers of the method with the high fan-in value and the relative position of the call in the body of the caller. This analysis investigates whether the call occurs on the first, second, first before last, or last position. These positions would typically indicate a before or after advice as a natural aspect-refactoring solution for the candidate seed and its set of callers.

## 4.  THE CASE STUDIES

We have applied fan-in analysis to several case studies, three of which we describe in detail in this paper. All cases are open source systems, allowing validation of our results by others. The PETSTORE and JHOTDRAW systems are demonstration applications of J2EE technologies and design patterns, respectively. TOMCAT is the largest system, and one that is widely used in web servers all over the world.

Before going into detail in the case studies, we first discuss a number of general observations, and explain in what format we will present the three case studies.

### 4.1   First Findings

Key statistics for our case studies are provided in Table I. A first observation that can be made from this table is that filtering methods above the threshold of 10 reduces the number of methods to be inspected to 1, 6, and 3 percent for PETSTORE, JHOTDRAW and TOMCAT, respectively. Figure 5 shows the fan-in distribution for the three case-studies. As can be seen, the vast majority of methods have a very low fan-in. The large percentage of methods with a fan-in value of 0 can be explained by the nature of the applications. PetStore, for instance, is a J2EE application and a number of calls are not explicit in the code but made by the container (EJB-specific methods).

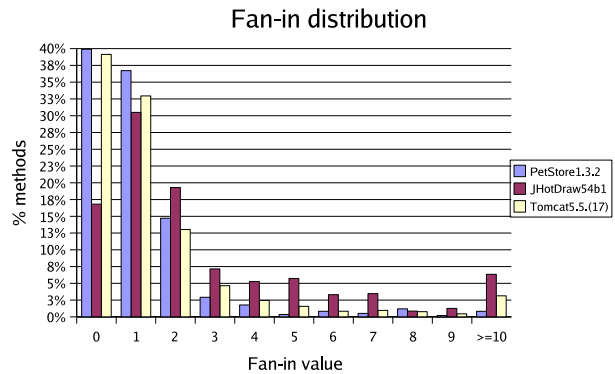

Fig. 5.    Fan-in distribution for the three case studies.

A second observation that can be made from Table I is that the accessor and utility filters eliminate about half of the high fan-in methods. Note that the utility methods filtered out here are the ones that are part of the system under study. Utility methods in external libraries are not taken into account in the first place, and do not occur in the table. If necessary, the scope of the system under study can be extended to include certain libraries as well. This is a decision that requires a certain amount of domain knowledge, for example that a particular library is used for addressing a known crosscutting concern (we will encounter such a situation for the *logging* concern in the TOMCAT case study in Section 7).

The methods of the system under study that are not filtered out will give the set to be analyzed in a last, tool-assisted step. This should result in a classification as either a seed for a crosscutting concern, or as a non-seed. Our third observation from Table I is that for all cases, a significant percentage (87%, 51%, and 73% for the three cases) of the methods that need to be inspected manually turn out to be confirmed seeds. Thus, while this step may be more labor-intensive, it does give a good chance of finding crosscutting concern seeds.

A final observation is that there are many more seeds than concerns. This is due to two reasons. First, there may be multiple concern instances for one sort of concern. For example, JHOTDRAW makes use of more than one Observer. Second, a single concern is often identified through multiple seeds. For example, for the Observer design pattern, we may not only find a high fan-in for the notification method, but also for the methods for attaching different observers to a subject.

### 4.2   Case Study Presentation

In the next sections we discuss the PETSTORE, JHOTDRAW, and TOMCAT case studies. We particularly focus on the third step, in which seeds are either confirmed or rejected as belonging to a crosscutting concern, since this step implies various considerations, inherent in the mining process, about the classification of a candidate seed. For each case study, we discuss several of the concerns found in considerable detail, explaining why we think that they are crosscutting, and analyzing to what extent these concerns are amenable to an aspect-oriented re-implementation. A full list of all high fan-in methods and the concerns they belong to are publicly available on our web site[4] as well as in the technical report on which this paper is based [Marin et al. 2004b]. The site furthermore describes which methods exactly were marked as utilities, thus making our experiments fully reproducible.

| Method | Fan-in | Concern |
|---|---|---|
| XMLDocumentException(String) | 27 | Contract enforcement |
| ServiceLocatorException(Exception) | 22 | Exception wrapping |
| CatalogDAOSysException(String) | 19 | Exception wrapping |
| PopulateException(String, Exception) | 11 | Exception wrapping |
| TransitionException(Exception) | 15 | Exception wrapping |
| XMLDocumentException(Exception) | 23 | Exception wrapping and tracing for debugging |
| ejb.ServiceLocator() | 30 | Service locator |
| XMLDBHandler() | 10 | *False positive* |

Table II. PETSTORE high fan-in methods and concerns

In order to give an impression of the limitations (and hence opportunities for improvement) of fan-in analysis, the next sections also discuss some of the false positives (rejected candidate seeds) and some of the concerns that are known from the literature or from related studies that our analysis missed (false negatives). Note that while we can compute the percentage of false positives (the number of non-seeds divided by the total number of seeds), we cannot determine the percentage of false negatives. This would require a common benchmark that documents all the crosscutting concerns exhibiting the symptoms (code scattering) targeted by fan-in analysis. At the time of writing, no such benchmark exists.

## 5. PETSTORE

The first case study we discuss is PETSTORE. This is a sample J2EE e-business application developed by SUN.[5] It is a demonstration of a Web application allowing customers to purchase via a web browser. In addition, it includes modules to perform administration tasks like sales statistics, orders and shipping management, etc. PETSTORE is an application demonstrating the proper use of most of the J2EE concepts, and can be considered a well-designed system.

An overview of the methods with a fan-in of 10 and higher, their fan-in value, and the concerns they represent is given in Table II. In this paper we explain why these concerns are indeed crosscutting. A detailed description of their refactoring towards ASPECTJ is presented by Mesbah and van Deursen [2005].

**Service Locators** The method with the highest fan-in value (30) belongs to the *Service-Locator* class from the *ejb* package, which implements the J2EE pattern of the same name [Alur et al. 2003]. The intent of the pattern is to provide a single point of control to clients that need to locate and access a component or service in the business or integration tier. The common solution is to have a single instance of the service locator class for an application or, at least, for a tier and thus to have it implemented as a singleton. The advantages of this solution, however, are not always clear for the EJB-tier and thus the adopted solution can vary [Johnson 2003].

---

[4] `http://swerl.tudelft.nl/bin/view/AMR/FanInAnalysisResults`
[5] `http://java.sun.com/blueprints/`, PETSTORE version 1.3.2.

```
public class InvoiceTD implements TransitionDelegate {

 /** sets up all the resources that will be needed to do
  *  a transition
  */
 public void setup() throws TransitionException {
  try {
    ServiceLocator sl = new ServiceLocator();
    qFactory = sl.getQueueConnectionFactory(JNDINames. ...);
    q = sl.getQueue(JNDINames. ...);
    queueHelper = new QueueHelper(qFactory, q);
  } catch(ServiceLocatorException se) {
   throw new TransitionException(se);
 }}

 /** Send an order approval to the OrderApproval Queue...
  */
 public void doTransition(TransitionInfo info) throws TransitionException {
  String xmlCompletedOrder = info.getXMLMessage();
  try {
    queueHelper.sendMessage(xmlCompletedOrder);
  } catch (JMSException je) {
   throw new TransitionException(je);
}}}
```

Fig. 6.   Error handling in PETSTORE

PETSTORE contains two different service locators: the web-tier one is implemented as a singleton but the fan-in of the method returning the unique instance is only 7; the identified EJB-tier locator is not a singleton and the method reported is the constructor of the class.

The service locator defines a consistent lookup mechanism for the dependencies of the various application components, which couples these components to the infrastructure framework and tangles them with the lookup logic.

A possible refactoring for the service locator is the *Dependency Injection* pattern (also called *Inversion of Control*) used in lightweight containers to avoid directly referencing a service locator [Fowler 2004], a mechanism that resembles the aspect-oriented mechanisms for injection. For Singleton implementations, the aspect refactoring of the pattern [Murali et al. 2004] and the optional caching mechanism [Laddad 2003b] are in place. The exception wrapping discussed next is also applicable to the Service Locator identified.

**Exception Wrapping**   The majority of the seeds are constructors for PETSTORE exceptions. As an example, Figure 6 shows the *TransitionException* case, which is thrown from 15 `catch` blocks in different classes and packages.

As in the *InvoiceTD* class in the figure, most of the methods throwing the exception implement `doTransition(..)` and `setup()` declared by the *TransitionDelegate* interface. All the transition delegates handle exceptions related to the particular functionality and re-throw *TransitionException*. This mechanism is common to many J2EE design patterns [Alur et al. 2003], such as *Business Delegate* discussed by Laddad [2003a]. The exception wrapping in *Business Delegate* aims at hiding the implementation details of a business service. The issue hidden in this case is the sort of exception that can be thrown by the actual implementation.

This consistent mechanism is spread over many places, and a refactoring solution is discussed by Laddad [2003a]. Aspects can be used to isolate the exception handling and to wrap the original exception thrown by the underlying implementation in the new exception. This will result in improvements in code size, localization and clarity. Studies of exception handling refactoring [Lippert and Lopes 2000] show a reduction of `catch` statements when using AOP of up to 95%. For the case at hand, we found that the classes affected were reduced by 20% [Mesbah and van Deursen 2005].

**Contract Enforcement**  A method with a fan-in value of 27 is a constructor for the *XML-DocumentException* class. This exception is raised if the structure of the XML document does not comply with the expected structure. By examining the call sites, we were able to observe that 9 of them are `fromDOM(Node)` methods, all throwing the exception if a certain compound condition fails. It turns out that all complex conditions share a common check, which can be easily factored out as an aspect by means of before advice – giving rise to the concerns similar to the pre- and post-condition (design by contract) examples discussed by The AspectJ Team [2003].

In this manner, the code will be better localized and new methods will be prevented from omitting the required checks.

Moreover, a set of another 14 call sites are methods of the same class that throw the reported exception if certain conditions do not hold. A sub-set of 11 methods from these callers check the same condition, namely the Boolean value of an input parameter.

**Debug Information**  The *XMLDocumentException* class has a second constructor with a high fan-in. This constructor is (like for the business delegates) used as an exception wrapper. In addition to that, before being wrapped the exception at hand is written on the error output stream. This additional behavior (on top of the wrapping) can be added as another aspect, which indicates which exception should be printed before being wrapped. Turning printing debug information into an aspect helps to ensure a common debugging strategy, and to isolate the concern that is otherwise crosscutting.

**False Positives**  The one case considered as non-aspect in the first set of candidates is an `XMLDBHandler` constructor with a fan-in value of 10. The callers are `setup(..)` methods in classes that populate the associated database tables with data from XML files. The `setup(..)` implementations are only slightly different: they return an instance of an anonymous inner class extending *XMLDBHandler* that is an XML filter. Because all the callers are well localized in a single package and there is only one `populate(..)` method that triggers the whole process at a client's request, we decided to label this candidate as non-crosscutting.

**False Negatives**  As briefly mentioned at the beginning of this section, one of the missed concerns is the service locator in the web-tier, implemented as a singleton, but whose method for accessing the unique instance has a fan-in value of only 7.

A second concern potentially identifiable by fan-in analysis is transaction management. If J2EE's built-in transaction mechanism is used, the concern is well-isolated. PETSTORE, however, also includes explicitly encoded transaction management, which consists of calls to the Java Transaction API (JTA). In principle these can be detected by fan-in analysis, but since they belong to an external library, we normally would not include them in our analysis. Furthermore, the fan-in values for the two methods in the JTA API (the

| Concern | No. of methods | Max fan-in |
|---------|----------------|------------|
| Adapter | 1 | 37 |
| Command | 2 | 24 |
| Composite | 12 | 24 |
| Consistent behavior | 20 | 31 |
| Contract enforcement | 3 | 31 |
| Decorator | 6 | 57 |
| Exception handling | 1 | 11 |
| Observer | 10 | 37 |
| Persistence | 6 | 22 |
| Undo | 3 | 25 |

Table III. Concerns found for JHOTDRAW, together with the number of high-fan in methods, and the highest fan-in among those methods.

*javax.transaction.\** package) used by PETSTORE code have a value (of just 2) well below our threshold.

## 6. JHOTDRAW

JHOTDRAW[6] is an application framework for two-dimensional graphics. It is an exercise in developing software making use of design patterns [Gamma et al. 1994].

Our filters eliminated around half of the methods with top fan-in values. We were rather cautious not to eliminate too many methods. The only methods designated as "utility" are enumeration manipulators (e.g., `FigureEnumerator.hasNextFigure()/next-Figure()`).

An overview of the concerns found is given in Table III. For each concern, it lists the number of different high fan-in methods that pointed to the concern, and the maximum fan-in value for this concern. In the next sections we discuss these concerns in more detail. Aspect solutions for some of these concerns are available through the open source AJHOT-DRAW[7] project [van Deursen et al. 2005], an ongoing activity to refactor JHOTDRAW to ASPECTJ starting from the results reported in the present paper.

### 6.1 The Undo Concern

In the top of the list of methods sorted by fan-in, a number of methods point to the undo functionality, such as the `undo` method in *UndoableAdapter*. An undo in a graphical editor is clearly a concern that cuts across many features and activities, although textbooks on aspect-oriented programming, such as Gradecki and Lesiecki [2003], The AspectJ Team [2003], Laddad [2003b], do not discuss using aspects for undo functionality .

A (somewhat simplified) representation of the participating classes in the JHOTDRAW undo implementation is given in Figure 7. JHOTDRAW offers various sorts of *activities*, which are contained in a class hierarchy. Examples of concrete activities include handling font sizes, triangle rotation, or image rotation.

The interface *Undoable* encapsulates the notion of undoing an action, for which it provides the `undo` method. Each class implementing a concrete activity that can be undone defines a static nested class conforming to this *Undoable* interface. The nested class knows how to undo the given activity, and has access to all the details of the activity that may be needed for this. Whenever the activity modifies its state, it also updates fields in its associated undo-activity needed to actually perform the undo. In addition to that, a list of *affected figures* is maintained, whose state must be adjusted if the activity is to be undone.

---

[6] `http://jhotdraw.org/`, version 5.4b1

[7] `http://ajhotdraw.sourceforge.net/`

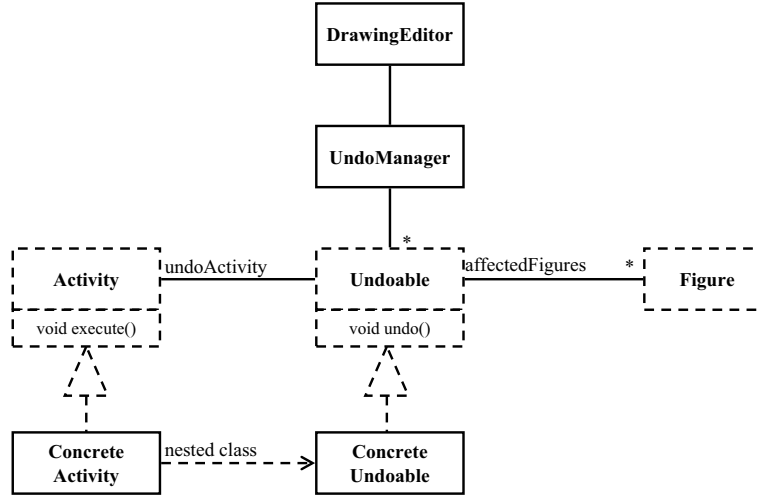

Fig. 7.   Participants for *undo* in JHOTDRAW.

In JHOTDRAW, there are 22 activities that can be undone, causing the undo concern to be spread over these classes. This, in turn, leads to a high fan-in for the methods of, for example, *Undoable*, which helped us to identify this crosscutting concern.

An aspect-oriented solution for the undo concern is presented by Marin [2004]. It consists of a number of steps.

—First, the existing activities are extended with an association to their undoables by means of an inter-type declaration.

—Second, existing operations are extended with functionality to keep track of the old state so that the action can be undone. These existing operations can be captured using a pointcut, and then the updates can be contained in advice code.

—Last but not least, the various nested classes containing the undoable activities can be added by means of inter-type declarations.[8]

Thus, this refactoring captures the undo "protocol" in a pointcut and advice, ensuring that undo functionality is properly invoked whenever commands are executed. Furthermore, the methods and (inner) classes devoted entirely to undo functionality are moved out of the command classes, and are remodularized into an aspect.

___________
[8] The present version of ASPECTJ does not support introducing inner and static nested classes.

## 6.2 Persistence

Another crosscutting concern that pops out clearly through a high fan-in is persistence. The concern was easily spotted, as there are six different methods involved, each having a name built from words like "read", "write", "storable", "input", and "output". Storing and restoring figures is performed by methods inherited from the *Storable* interface. This interface offers methods to read one self from a *StorableInput* stream, or write one self to a *StorableOutput* stream.

The implementation of the persistence concern is spread over 36 classes. Figures implementing the *Storable* interface invoke several methods from the *StorableOutput* and *StorableInput* classes. The two classes are specialized in writing/reading various (primitive) types, (e.g., String, Color, int, etc.) to/from a storing device. This results in a high fan-in for their methods, which allowed us to detect the persistence concern using fan-in analysis.

The *Storable* interface can be considered a *secondary* interface, i.e., one that does not define the primary role of the implementing class but only adds supplementary functionality to it. An aspect-oriented implementation for this concern can super-impose such as secondary role onto relevant classes by means of inter-type declarations (as done in the AJHOTDRAW project [Marin et al. 2005]). In this way, the persistence logic is isolated in the aspect, and figure classes need not contain any persistence-related code.

Observe that this refactoring merely moves methods from classes to aspects, and involves neither a pointcut nor advice. Thus, this refactoring does not have an effect on any *fan-in* value, and the methods from the *StorableOutput* and *StorableInput* classes will continue to have a high fan-in. In the original implementation, however, these calls came from many different classes or even different packages. In the aspect solution, all calls are from the persistence aspect. This suggests that it may be interesting to lift the call relation to the class, aspect, or package level, and count, for example, the number of other packages using a particular method. We have not yet explored this direction.

## 6.3 Observers in JHOTDRAW

Several methods with high fan-in point to instances of the *Observer* design pattern. Example methods include `Figure.addFigureChangeListener(..)` (fan-in 11) and `Figure.changed()` (fan-in 36).

The participants of the Observer design pattern are shown in Figure 8, taken from [Gamma et al. 1994]. One method that we expect to have a high fan-in is `notify`: this method is called for every different kind of change event the observer wants to hear about. Furthermore, we expect the fan-in for the `attach` and `detach` methods to be related to the number of observers involved. The `Observer.update()` method is likely to have a low fan-in value, as it is only called from the `Subject.notify()` method.

These expectations are met in JHOTDRAW: The `Figure.changed()` method corresponds to the `Subject.notify()` and indeed has the highest fan-in, allowing us to discover this concern. Observers are called *Listeners* in JHOTDRAW, and the `addFigureChangeListener` corresponds to the `attach` method.

Matching on the naming conventions used in the first observer found led us to another instance of the pattern (with a somewhat lower fan-in). Thus, fan-in analysis provides initial seeds and application understanding, which then can be used by complementary techniques to identify further cross cutting concerns.

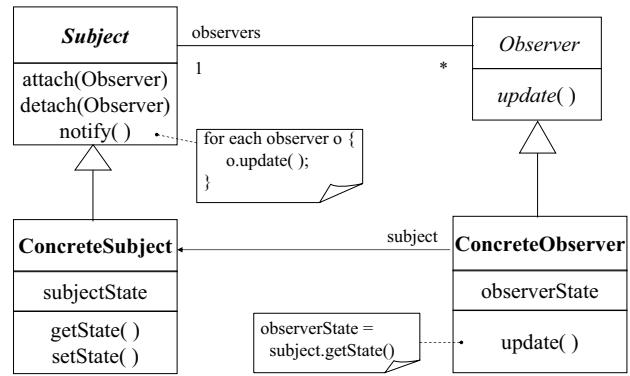

Fig. 8.   Class diagram illustrating the participants in the Observer design pattern.

```
public void execute() {
  // perform check whether view() isn't null.
  super.execute();

  // prepare for undo
  setUndoActivity(createUndoActivity());
  getUndoActivity().setAffectedFigures(view().selection());

  // key logic: cut == copy + delete.
  copyFigures(view().selection(), view().selectionCount());
  deleteFigures(view().selection());

  // refresh view if necessary.
  view().checkDamage();
}
```

Fig. 9.   (Simplified) execute method in JHOTDRAW exhibiting tangling.

The Observer is a prototypical example of a design suitable for an aspect implementation: Inter-type declarations can be used to super-impose the *Observer* or *Subject* roles onto classes of interest, and pointcuts and advice can be used to weave in the appropriate calls to `notify()`.

The notification protocol used in JHOTDRAW is somewhat more complicated than a simple call to `changed()`. Before the change is being made, the affected figures should be invalidated, which should be done by means of a call to the method `willChange()` (fan-in value 25). Such a *policy enforcement* concern calls for an around advice, which helps to ensure that the protocol is properly implemented.

### 6.4 Other Concerns

**Command and Related Concerns**  A method with high fan-in value (24) that is easy to connect to a design pattern is `AbstractCommand.execute()`. The crosscutting nature of the Command pattern is discussed by Hannemann and Kiczales [2002]. They propose a (fairly complex) aspect-oriented representation in which different roles (such as the command *invoker* and *receiver*) are distinguished. The *protocol* between these is based on a pointcut capturing all places where invocations are required (for example when a GUI button is pressed). The advice then is to activate the receiver for the given invoker. This corresponds to calling the *execute* method, which in the aspect solution has a low fan-in, and in the non-aspect implementation a high one. The applicability of this solution to JHOTDRAW is not clear: isolating the Command concern in this way is complicated by the interaction with the *undo* and *redo* concerns.

The various implementations of the specific `execute()` commands exhibit two further concerns, as illustrated by the *CutCommand* example in Figure 9:

—Each `execute` implementation starts with a super call responsible for checking a common pre-condition, throwing an exception if it does not hold. This is a *Contract enforcement* concern as discussed for PETSTORE.

—Most `execute` implementations conclude with a check if the figure has been changed in order to trigger a refresh of the view if necessary. This is a *Providing consistent behavior* concern as discussed by The AspectJ Team [2003].

Factoring these (as well as the undo functionality) out of the code in Figure 9 would leave the `execute` method with just its core functionality, which is an implementation of the cut operation by means of a copy and delete operation.

**Consistent Behavior**  The seeds reported by fan-in analysis cover 11 different instances of the "consistent behavior" concern. In other words, there are 11 different contexts into which a set of method-callers invoke a method with a high fan-in value as part of a consistent mechanism. Examples include the previously discussed notification to conclude the execution of commands, consistent (de-)activation of tools, initialization of tools, etc. Each of these 11 instances is a suitable candidate for replacement by an aspect solution by means of a pointcut and advice.

**Composite**  High fan-in values are also obtained for the children manipulation methods from the Composite pattern (e.g., `add(Figure)`, fan-in value 13). The high fan-in in this case is largely due to the fact that these manipulation methods are widely used, but there was no systematic pattern in this usage. The high fan-in is not directly related to the crosscutting nature of the Composite pattern, and, consequently, not affected by a refactoring to the aspect-oriented Composite implementation suggested by Hannemann and Kiczales [2002] (which consists of one aspect containing inter-type declarations for the various composite participants).

**Decorator, Adapter**  Several of the high fan-in methods are related to the Decorator or Adapter patterns. These patterns are different from, e.g., Command and Observer, which have characteristic methods likely to have a high fan-in (execute and notify, respectively). Instead, the Decorator and Adapter patterns make use of consistent forwarding, which allows us to recognize the relation with the pattern of the several methods with a high fan-in value reported for this concern (such as `DecoratorFigure.containsPoint`, fan-in value 15).

The aspect solution for these patterns as discussed by Hannemann and Kiczales [2002] is to drop the decorator and adapter classes altogether, directly weaving in the relevant decorations or adaptations in the appropriate classes. Whether this solution is applicable to JHOTDRAW is not clear, since JHOTDRAW relies on enabling or disabling decorations (which is less easy to do in the implicit aspect solution).

**False Positives**  The group of false alarms for JHOTDRAW consists of 56 methods. More than half of these methods are implementations of two methods: `displayBox` and `containsPoint`. The first of the two returns the display box of a figure. The method has a high fan-in value because it supports many of the actions associated with a figure, like drawing or moving figures, etc. However, the callers could not be grouped by a clear relationship, and no clear call idiom could be observed when investigating the call sites.

Similar observations apply to the `containsPoint` method, which checks if a point is inside a figure. Except one implementation, which together with other reported methods in the *DecoratorFigure* class implement the consistent logic of redirecting incoming calls, *containsPoint* has been marked as a false positive.

Other false alarms include five `moveBy` methods from *Figure* classes, which implement actions for moving a figure, and a number of complex accessor methods that could not be filtered using the name or implementation criteria.

**False Negatives**  As discussed for the identified Observer pattern instance, other instances of this pattern can be discovered starting from the fan-in seeds. The *Drawing* classes, for example, are part of a different Observer implementation and define role-specific methods with names that are similar to those in the *Figure* classes: `add/removeDrawingChange-Listener(..)`. These role methods have lower fan-in values because the *Drawing* Observer implementation has a smaller extent, with fewer classes that register as *Drawing* observers.

The comparison experiment using JHOTDRAW as common benchmark revealed a few concerns missed by fan-in analysis [Ceccato et al. 2006]. One of these concerns is a *Visitor* pattern instance. The pattern defines specific roles and methods, such as the visit operations for the *Visitor* role, and the `accept` method implemented by the *Visitable* elements. The `visit` method in the *Visitor* role would collect calls from all the *Visitable* classes that pass self-objects as arguments to this method for being visited. A large number of *Visitable* elements would therefore increase the fan-in value of the visitor method. However, in JHOTDRAW only two *Figure* classes implement the methods to accept visitors. The large majority of figures do not override the default implementation for this task, which also implements the tree traversal for composite elements.

We have found implementation of the Visitor pattern through the role-specific methods by applying FINT to its own source code, as well as in TOMCAT, as we shall see in the next section.

## 7.  TOMCAT

Apache TOMCAT is the servlet container that is used in the reference implementation for Sun's Java Servlet and JavaServer Pages technologies. TOMCAT is developed within the

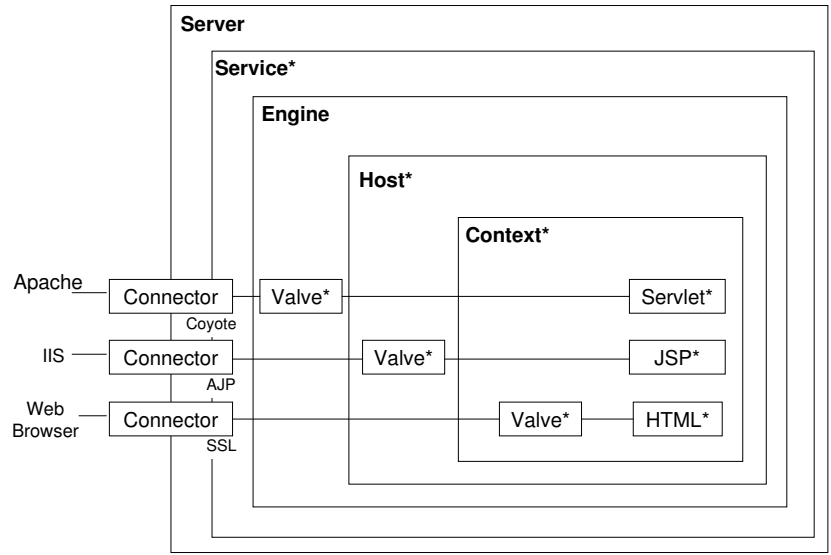

Fig. 10.   Example TOMCAT configuration

open-source Jakarta project at the Apache Software Foundation.[9]  The main elements of
TOMCAT are the servlet container called Catalina, the JSP engine called Jasper, and the
TOMCAT connectors.  We analyze and discuss the results for version 5.5(.17) of TOM-
CAT [10].

The main architectural components of TOMCAT are shown in Figure 10 [Moodie 2005].
The outer *Server* component offers a number of *Services* through various *Connectors*. The
default connector implements HTTP. The *Engine*, *Host* and *Context* components are all
*container components*, representing the top-level container, the virtual host, and the actual
web application, respectively. Inside containers there can be *nested components* which can
provide various administrative services. Some components can be contained more than
once and are marked with a star in the figure. Particularly relevant for our discussion are
the nested components called *Valves*: these can intercept a request and process it before it
reaches its destination.

The crosscutting concerns found for TOMCAT are summarized in Table IV. Again, some
of the concerns are related to crosscutting behavior as encountered in design patterns, but
there are also some concerns not previously described. Below we elaborate some of the
concerns in more detail.

### 7.1  Lifecycle

*Lifecycle* is a common interface for several Catalina components, providing a consistent
mechanism to start and stop the component. It is a secondary interface, adding new, sup-
plementary capabilities to the core logic of the implementing classes. *Lifecycle* is imple-mented by more than 40 classes. The `start` and `stop` methods for these classes have fan-in values varying between 25 and 34. The set of results of fan-in analysis comprises 73 implementations of these two Lifecycle methods.

---

[9] `http://jakarta.apache.org/tomcat/`
[10] `http://tomcat.apache.org/tomcat-5.5-doc`

| Concern | No. of methods | Max fan-in |
|---|---|---|
| Chain of responsibility (pipeline) | 24 | 18 |
| Command | 2 | 16 |
| Composite | 9 | 37 |
| Consistent behavior | 34 | 90 |
| Contract enforcement | 9 | 46 |
| Lifecycle | 73 | 34 |
| Logging | 1 | 10 |
| Observer | 6 | 56 |
| Redirector | 4 | 25 |
| Visitor | 1 | 28 |

Table IV. Concerns found for TOMCAT, together with the number of high-fan in methods, and the highest fan-in among those methods.

The `start` and `stop` methods are part of a particular type of *consistent behavior* scheme: The `start` operation has to be called before any public method of the component, while `stop` terminates the object's use and should be the last call for a component's instance. Furthermore, implementors of the *Lifecycle* interface have to adopt the *Subject* role from the *Observer* pattern: listeners can be added which must be notified of start or stop events. The key methods to support these operations have fan-in values as high as 56.

The Lifecycle concern can be seen as a generalization of the use of `stop()` methods to remedy Java's expensive finalization mechanism [Vickers 2002; Goetz 2004]. Those methods take care of cleaning up the object's resources inside the program code to avoid the overhead of having finalizers but will result in crosscutting for the object's clients.

The Lifecycle concern is complex, comprising several crosscutting concerns. Although aspect-oriented solutions have been presented for some parts of it, a complete refactoring solution remains an open issue. One of the problems is that the type of consistent behavior needed by the concern cannot be expressed in a pointcut-based aspect language like AS-PECTJ (because it requires specifying "before accessing any public methods of class" and "after last use of class").

### 7.2 Valves / Chain of Responsibility

A method occurring around 20 times in the seed list is the `invoke(..)` method in the *Valve* hierarchy. Valves are nested components that implement a pluggable request-processing operation for an associated container. Valves are connected through a pipeline structure, in which each valve passes the request to the `invoke` method of the next valve in the pipeline. Examples of valve classes include *AccessLog Valve* to create standard web servers log files, *RemoteAddress Valve* to filter the requests by the IP address of the client that submitted them, or *SingleSignOn Valve* to grant user access to the web applications associated with a virtual host.

The pipeline organization of the valves is implemented using the *Chain of responsibility* pattern [Gamma et al. 1994]. This implies that a valve's core logic is crosscut by the functionality of retaining the reference to the next valve in the pipeline and consistently passing the invocation to it. Furthermore, the various implementations of the `invoke` method are tangled with other concerns. The *AuthenticatorBase* abstract class, for instance, implements the basic functionality of the request authentication valve. However, its `invoke` method also performs logging operations for debugging activities. Similarly, the previously mentioned *AccessLog Valve* implements a *timing* operation for the request/response operation it has to log. An aspect-oriented solution for the *Chain of responsibility* pattern is provided by Hannemann and Kiczales [2002].

### 7.3　Other Concerns

A number of architectural components of TOMCAT and Catalina are Container elements. The *Container* interface defines these elements as *Composite* structures. Standard implementations of the interface are abstractions of the TOMCAT container components, like *StandardEngine* or *StandardContext*. Fan-in analysis identifies the children manipulation methods specific to the *Composite* structure of these components and reports them as concern seeds (fan-in values of up to 37).

In the same category of design patterns, a number of seeds correspond to the Observer, like the notifier for Container events (`ContainerBase.fireContainerEvent(..)`) (fan-in value 55) and the `execute` method of the *Command* pattern implementation (fan-in value 16). Similar to the cases discussed for JHOTDRAW, the Command seed methods reported for TOMCAT are also part of a *contract enforcement* that consist of a pre-execution attribute validation. The contract is implemented as a call to the method in the super class. Other seed results include methods that participate in the implementation of consistent *redirection* functionality (*Wrappers*); the methods implement non-trivial accessors that are invoked by a large number of methods that simply redirect their callers to dedicated methods of the reference returned by the reported seed. The fan-in values for these seeds are up to 25.

Different *pre-condition check enforcements* are also part of the various implementations for the *Lifecycle* `start` and `stop` methods. The reported seed method in this case is the constructor of the exception thrown if the pre-condition does not hold (fan-in value 32).

The *logging* concern is particularly interesting because of the new implementation strategy in version 5.x of TOMCAT. This concern used to be implemented in the previous versions using Logger classes that were part of the Catalina API. However, the current implementation uses logging functionality available through specialized, external libraries. Although we have been able to directly identify *logging* methods in the analyzed code (e.g., *ModuleClassLoader*), as well as logging functionality tangled with the implementation of other seed methods, a number of direct *logging* seeds are missed. This is due to our choice not to include library components in the analysis, as discussed in Section 4.

The remaining seeds include, besides other instances of the concerns already discussed, a large number (up to 25) of different instances of the *consistent behavior* concern, as well as seeds for the super-imposed role in the *Visitor* pattern.

**False Positives**　A group of 13 false alarms consists of methods in the *JspReader* and *ServletWriter* classes. The first class is an input buffer for the JSP parser, and the reported methods are utilities for parsing JSP files, like methods to match an input String in a file or to skip space-characters. The callers are methods in the JSP *Parser* class.

The methods reported for *ServletWriter* print String elements in various formats to an output stream. The callers of these methods belong to the *Generator* class, which outputs Java code from an internal, tree-based (XML) representation of JSP documents.

These classes could have been considered as *utility*, if we would have had more detailed knowledge about the system prior to analysis.

Among the other false alarms there are 12 implementations of the `store` method in the *StoreFactoryBase* hierarchy. The classes in this hierarchy are specialized in storing configuration elements, such as *Server*, *Service*, *Engine*, or *Context* to a XML configuration file (server.xml). The callers of the reported methods are declared in classes in the same hierarchy or are overloaded implementations of the `store` method in the class *StoreConfig*. This class is part of the same concern as the reported methods and so no crosscutting element could be identified.

**False Negatives** The literature on TOMCAT discusses hardly any crosscutting concerns, making it difficult for us to assess whether there are any interesting false negatives we missed. The crosscutting concern that is discussed widely for TOMCAT is logging, and often it is mentioned as an example of poor modularization. As already discussed, fan-in analysis helps us to identify several seeds for the logging concern. However, the analyzed version of TOMCAT is extensively using logging methods declared by external libraries (the *org.apache.commons.logging.\** package). By canceling the filter for library methods in FINT and looking for calls to externally declared methods, we noticed that there are 19 methods from the logging package that are referred from the analyzed TOMCAT sources. From these ones, 13 methods belong to the *Log* class and show a fan-in value higher than the considered threshold of 10. The fan-in value for the logging method for debugging (`Log.debug`), for example, is as high as 465.

## 8. DISCUSSION

**High Fan-in as Indicator** As we have seen in the previous case studies, fan-in analysis identifies high fan-in methods, applies a series of filters to these methods, after which more than half of the remaining methods turn out to be related to a crosscutting concern.

We can distinguish three main situations in which a high fan-in value indicates the presence of crosscutting concerns:

—The method has a high fan-in because it is part of a *dynamic* crosscutting mechanism. The typical refactoring will be to capture the call sites through a pointcut, and to move the method call to advice. Examples that we encountered include exception wrapping, contract enforcement, observer notification, and life cycle.

—The method has a high fan-in because it is used by a *static* crosscutting mechanism. A typical example is a secondary interface that must be implemented by a series of classes. The various implementations are likely to make use of the same helper methods, giving these a high fan-in. The refactoring is to collect all these interface implementations into one or more inter-type declarations. This we encountered for the persistence concern.

—The method has a high fan-in because it is part of a concern that plays a key role in the design. The method happens to be part of a crosscutting concern, which will benefit from an aspect-oriented refactoring. The refactoring, however, will not affect any of the call sites of the high fan-in method. This we encountered for the composite concern.

These situations are not mutually exclusive. In many cases, a concern involves static as well as dynamic crosscutting, as we have seen for the undo concern. We then are likely to see multiple seeds, which may either point us to the static or to the dynamic crosscutting behavior.

**The Type of Concerns Identified**   The fact that we were able to find similar aspects in various case studies suggests that their identification is not accidental. We identified various crosscutting concerns that are discussed in the literature, including those that stood at the origins of aspect-oriented programming. In addition, we have identified a number of new aspects, such as *Undo* and *Lifecycle*. Given the different nature of the three case studies, we feel that these results can also be achieved for other cases.

A notable source of crosscutting behavior is formed by various design patterns: for both JHOTDRAW as well as TOMCAT they account for approximately half of the concerns identified. This suggests that it may be worthwhile to investigate the use of design pattern mining techniques (see, e.g., Ferenc et al. [2005]) for aspect mining purposes.

**Reasoning about Seeds and Non-seeds**   One of the subjective elements of our aspect mining approach is the third step in which the human engineer has to distinguish seeds from non-seeds. We adopted the following reasons for classifying a high fan-in method as a seed:

—We were able to link the method to a concern that is known to be crosscutting.

—We considered the method's concern to be conceptually separate from the key functionality of the calling classes. Thus, it would be meaningful to make the base implementation oblivious of method's concern.

—We could discover an idiom, recurring patterns, or other similarities in for example the call sites found, suggesting an implicit relationship between these call sites that could be made explicit through a pointcut with advice.

—We were able to identify a refactoring to ASPECTJ that may be beneficial in terms of modularization, flexibility, or evolution. Usually, these refactorings were composed from basic refactorings as included in the catalogs provided by Laddad [2003a] and Monteiro [2004].

When rejecting a high fan-in method as a seed, we were not able to achieve any of the above.

**Utility Filtering**   A step requiring some manual effort is the filtering of utility methods. The intent of this is to remove groups of methods for which it is a priori obvious that they do not belong to crosscutting concerns. It is not necessary to capture all utility methods. Therefore, the amount of effort involved in this step is very limited: if it is not immediately clear if something is a utility, it is simply safe not to filter the method, and analyze it in detail if it turns out to have a high fan-in.

**Percentage of False Positives**   The percentage of false positives in the three case studies is 13%, 49% and 27% for PETSTORE, JHOTDRAW, and TOMCAT, respectively (see Table I). Based on these figures, and based on experiments we conducted with other systems, we conjecture that 50-75% of the candidate seeds that we identify automatically can be confirmed as belonging to a crosscutting concern.

Note that this percentage is conservative in two ways: First, we only discarded classes or methods as utilities when this was immediately obvious. Second, we only confirmed seeds when we clearly could see the crosscutting nature of the underlying concern. In other words, it is possible that with a more involved analysis of some of the *non seeds* from Table I these could turn out to be crosscutting concerns as well. For this reason, it is reasonable to expect that other systems will exhibit a similar (or perhaps higher) success rate.

**False Negatives**  While working on our case studies, analyzing their design and implementation in considerable depth, we did encounter several crosscutting concerns not found through fan-in analysis, some of which were discussed in the previous sections. As an example, for PETSTORE we found transaction management, scattered implementations of the Serializable interface, and opportunities for making use of ASPECTJ's approach to imitating multiple inheritance [Mesbah and van Deursen 2005]. As for the logging example discussed for TOMCAT, key methods implementing transaction management are likely to be missed as well, because they are part of imported libraries that we do not include in our analysis. Crosscutting concerns found in JHOTDRAW through other aspect mining approaches are discussed and compared by Ceccato et al. [2006]. An example concern fan-in analysis did not find is bringing a figure to the front or sending it to the back, simply because the methods involved were not called sufficiently often.

Based on these observations, we can make the following more general claims about the sort of crosscutting concerns that will not be found through fan-in analysis. First, the "footprint" of the concern should be above the threshold. Thus, if the concern involves dynamic crosscutting, the number of scattered calls should be higher than the threshold. Furthermore, if the crosscutting is purely static, the concern will usually not be found, unless the scattered implementation relies on shared functionality, and the number of call sites is higher than the threshold.

Note that the effect of the threshold is twofold. First of all, it helps us reduce the number of methods to be inspected. In addition to that, it allows us to find those aspects that are likely to significantly influence the modularity of the source code. Thus, while we certainly miss some crosscutting concerns, we are likely to find the ones that are most scattered, and hence good candidates for refactoring.

**Percentage of False Negatives**  How to arrive at a percentage for false negatives is less clear. This would require a report of all the crosscutting concerns that could be found in the case studies considered. Such reports have not been available prior to our experiments. Furthermore, such a report would be affected by the difficulty of deciding objectively what is and what is not a crosscutting concern.

The way to achieve progress in this direction is by establishing a common benchmark of known crosscutting concerns in existing systems. Such a benchmark would not only be a simple list, but also a summary of the reasons why certain concerns are deemed crosscutting. Our coverage of the concerns we found through fan-in analysis is aimed at establishing and promoting such a benchmark.

**Seed Inspection Effort**  How much effort is involved in inspecting seeds by hand? An important observation to make is that in many cases it is possible to decide for a group of methods together whether they constitute a seed. One reason for this is our treatment of polymorphism. In our definition of the fan-in metric one call could increase the metric value for several methods in the hierarchy of the invoked callee. Therefore, method implementations in the same hierarchy, which most commonly implement the same concern, also share many of their callers.

Such situations are very common in the cases analyzed. In TOMCAT, for instance, the over 200 seed and non-seed methods are implementations or declarations of only less than 100 distinct methods. As another example, the set of candidates for JHOTDRAW includes more than 20 implementations of the `displayBox` method, which we marked as non-seed. Grouping methods by their declarations as supported by FINT considerably reduces the investigation effort required for each method.

FINT offers further ways to reduce the manual effort involved in seed inspection. This includes various analyses to detect relations between the callers of a reported method with a high fan-in value, as discussed in Section 3.4. For example, by examining the callers (of any of the around 20 reported implementations) of the `invoke` method in TOMCAT's *Valves pipeline* concern, FINT shows that more than 80% of these callers are also `invoke` methods in *Valve* classes. The tool groups these callers as shown in Figure 4. Such relations are present for a significant number of discovered seeds, including crosscutting elements discussed for JHOTDRAW's *Undo* concern and the concerns in the *Command* hierarchy, as well as *Exception wrapping* concerns in PETSTORE.

**Required Expertise Level**   How much domain knowledge or expertise is required for conducting fan-in analysis? For the bottom-up approach, when we look for consistent invocations of the method with a high fan-in value from call sites that could be captured by a pointcut definition, little specific knowledge is needed. For the top-down approach more a priori knowledge is required. The top-down approach relies on easily observable relations between tool-reported candidates and known examples of crosscutting functionality; design patterns are the most common in our cases. The rules we employed for associating patterns to candidates are simple: the methods are part of the roles defining the design patterns and/or they execute actions specific to responsibilities of participants in the pattern implementation (e.g., delegations of actions).

Note, however, that many crosscutting concerns described in the context of design pattern implementations will typically be found by means of the bottom-up approach as well. For instance, calls to notification methods in implementations of the Observer pattern, or invocations to the action of the next element in a pipeline (chain of responsibility) are typical examples of crosscutting concerns targeted by fan-in analysis. In this case, the discussion of the patterns serves to describe the larger context into which the crosscutting concern occurs.

**AspectJ**   Fan-in analysis is an aspect mining approach that is entirely independent of ASPECTJ or any other aspect-oriented language. Fan-in analysis is a technique for understanding a system's modularization, helping developers to find crosscutting concerns. Some of these can be candidates for a refactoring towards ASPECTJ (as discussed for PETSTORE and JHOTDRAW by Mesbah and van Deursen [2005] and Marin et al. [2005]). For other concerns, alternative aspect-oriented solutions, such as composition filters [Bergmans and Aksit 2001] or the inversion of control pattern [Fowler 2004], while for still other concerns present aspect-oriented languages do not offer a suitable modularization mechanism yet.

Thus, fan-in analysis is not only a possible first step in refactoring to aspects. It also is a program comprehension technique that can help to understand crosscutting concerns in existing applications.

**The Fan-In Metric**   The variant of the fan-in metric we have used, has been optimized for aspect mining purposes, and, as shown in this paper, has brought us good results. An open question is whether this metric can be further improved. One possible route would be to lift the fan-in metric to the class, inheritance hierarchy, or package level, as we briefly discussed for the persistence concern of JHOTDRAW in Section 6.2. Fine tuning the metric such that it reflects, e.g., call site locations instead of the mere number of methods containing call sites is an issue for further research.

## 9. CONCLUDING REMARKS

### 9.1 Contributions

We consider the following as our three key contributions.

First of all, we propose a new, metrics-based, aspect mining approach. The approach aims at capturing crosscutting concerns by focusing on methods that are called from many places, and hence have a high fan-in. Our case studies show that after appropriate filtering more than 50% of these methods turn out to belong to a crosscutting concern.

Our second contribution is FINT, a tool that is freely downloadable that supports fan-in analysis. FINT not only shows how the fan-in metric and the filters can be implemented, but also offers support for the final manual step consisting of exploring the high fan-in methods and their call sites, and managing the seed-methods.

The third contribution consists of the extensive case studies we conducted. We argue in detail why we think that certain concerns are crosscutting in three existing open source Java systems. Some of these concerns were not previously described in the literature as crosscutting (such as undo or lifecycle). Moreover, in most cases we discuss alternative aspect-oriented implementations of these concerns. The resulting list of concerns and their manifestation in the three systems is relevant not only for fan-in analysis: it is of value for the validation of any aspect mining approach.

In addition to that, we offer an explanation of our results by identifying the factors contributing to the success of fan-in analysis as an aspect mining approach, as well as the limitations of the approach.

### 9.2 Future Work

We are presently in the process of extending our results along the following lines.

First, we are considering various extensions to FINT. One route is to integrate FINT with other concern elaboration tools, such as FEAT [Robillard and Murphy 2002] or the Concern Manipulation Environment CME [Harrison et al. 2004]. We could use such tools to explore and describe a concern or feature to its full extent, starting from the (partial) set of elements and relations identified by FINT as part of the crosscutting concern implementation.

Another option is to combine FINT with other automated aspect identification techniques, such as, for example, techniques based on formal concept analysis, identifier analysis, or clone detection. A prerequisite for combination is to be able to assess and compare aspect mining techniques and their results.

In addition to that, we continue to elaborate our case studies. This will provide further data on optimal threshold values, typical number of concerns that can be found in existing applications, and figures for the percentages of false positives and false negatives.

The results presented in this paper show that the recognized crosscutting concerns follow various implementation idioms. Fan-in analysis is particularly suited for identifying method invocations that cut across a set of other methods. However, concerns like those encountered in the Decorator pattern are typically less likely to occur among the results of this technique. Steps towards design of mining techniques that target specific implementation idioms have been taken in our more recent work [Marin et al. 2006]. Two such techniques are already available in FINT and more extensions are planned as future work.

One of our activities directly related to one of the case studies presented in this paper is AJHotDraw [van Deursen et al. 2005], a sourceforge project in which we offer an aspect-oriented re-implementation of JHotDraw, based on the concerns found in the present paper. In this way, the case studies presented here form the starting point for a benchmark for comparing aspect mining and refactoring approaches.

**Acknowledgments**   We would like to thank the anonymous reviewers for their feedback on earlier versions of this paper. The authors received partial support from SenterNovem, project Single Page Computer Interaction (SPCI).

REFERENCES

Alur, D., Crupi, J., and Malks, D. 2003. *Core J2EE Patterns*. Sun Microsystems, Inc., USA.

The AspectJ Team. 2003. *The AspectJ Programming Guide*. Palo Alto Research Center. Version 1.2.

Bergmans, L. and Aksit, M. 2001. Composing crosscutting concerns using composition filters. *Commun. ACM 44*, 10, 51–57.

Biggerstaff, T. J., Mitbander, B. G., and Webster, D. E. 1994. Program understanding and the concept assignment problem. *Communications of the ACM 37*, 5, 72–82.

Binkley, D., Ceccato, M., Harman, M., Ricca, F., and Tonella, P. 2005. Automated refactoring of object oriented code into aspects. In *Proceedings International Conference on Software Maintenance (ICSM 2005)*. IEEE Computer Society, Los Alamitos.

Breu, S. and Krinke, J. 2004. Aspect mining using event traces. In *International Conference on Automated Software Engineering (ASE 2004)*. IEEE Computer Society, Los Alamitos, CA.

Breu, S. and Zimmermann, T. 2006. Mining aspects from version history. In *ASE '06: Proceedings of the 21st IEEE International Conference on Automated Software Engineering (ASE'06)*. IEEE Computer Society, Washington, DC, USA, 221–230.

Briand, L. C., Daly, J. W., and Wüst, J. K. 1999. A unified framework for coupling measurement in object-oriented systems. *IEEE Transactions on Software Engineering 25*, 1, 91–121.

Bruntink, M., van Deursen, A., van Engelen, R., and Tourwé, T. 2005. On the use of clone detection for identifying cross cutting concern code. *IEEE Transactions on Software Engineering 31*, 10, 804–818.

Ceccato, M., Marin, M., Mens, K., Moonen, L., Tonella, P., and Tourwé, T. 2006. Applying and combining three different aspect mining techniques. *Software Quality Journal 14*, 3, 209–231.

Colyer, A., Clement, A., Harley, G., and Webster, M. 2005. *Eclipse AspectJ*. Pearson Education, Inc., NJ.

van Deursen, A., Marin, M., and Moonen, L. 2005. AJHotDraw: A showcase for refactoring to aspects. In *Proceedings of the AOSD Workshop on Linking Aspects and Evolution (LATE05)*. CWI, Amsterdam, The Netherlands.

van Deursen, A., Quilici, A., and Woods, S. 2000. Program plan recognition for year 2000 tools. *Science of Computer Programming 36*, 303–324.

Eick, S. G., Steffen, J. L., and Eric E. Sumner, J. 1992. Seesoft-A Tool for Visualizing Line Oriented Software Statistics. *IEEE Trans. Softw. Eng. 18*, 11, 957–968.

Ferenc, R., Beszédes, Á., Fulop, L., and Lele, J. 2005. Design pattern mining enhanced by machine learning. In *Proceedings International Conference on Software Maintenance (ICSM 2005)*. IEEE Computer Society, Los Alamitos, 295–304.

Fowler, M. 2004. Inversion of Control Containers and the Dependency Injection pattern. `http://martinfowler.com/articles/injection.html`.

Fowler, M. et al. 1999. *Refactoring: Improving the Design of Existing Code*. Addison-Wesley, Reading, MA.

Gamma, E., Helm, R., Johnson, R., and Vlissides, J. 1994. *Design Patterns: Elements of Reusable Object-Oriented Software*. Addison-Wesley, Reading, MA.

Goetz, B. 2004. Garbage collection and performance. IBM developersWorks articles. `www-136.ibm.com/developerworks/java/`.

GRADECKI, J. D. AND LESIECKI, N. 2003. *Mastering AspectJ - Aspect Oriented Programming in Java*. Wiley Publishing, Inc., Indianapolis, Indiana.

GRISWOLD, W. G., YUAN, J. J., AND KATO, Y. 2001. Exploiting the map metaphor in a tool for software evolution. In *ICSE '01: Proceedings of the 23rd International Conference on Software Engineering*. IEEE Computer Society, Washington, DC, USA, 265–274.

GYBELS, K. AND KELLENS, A. 2005. Experiences with identifying aspects in smalltalk using unique methods. In *Proceedings AOSD Workshop on Linking Aspect Technology and Evolution (LATE)*. CWI, Amsterdam, The Netherlands.

HANNEMANN, J. AND KICZALES, G. 2001. Overcoming the prevalent decomposition of legacy code. In *Proceedings of the ICSE Workshop on Advanced Separation of Concerns*. IBM Research, New York, USA.

HANNEMANN, J. AND KICZALES, G. 2002. Design pattern implementation in Java and AspectJ. In *Proceedings of the 17th Annual ACM conference on Object-Oriented Programming, Systems, Languages, and Applications (OOPSLA)*. ACM Press, Boston, MA, 161–173.

HARRISON, W., OSSHER, H., JR., S. M. S., AND TARR, P. 2004. Concern modeling in the concern manipulation environment. In *IBM Research Report RC23344*. IBM Thomas J. Watson Research Center, Yorktown Heights, NY.

HENDERSON-SELLERS, B., CONSTANTINE, L. L., AND GRAHAM, I. M. 1996. Coupling and cohesion (towards a valid metrics suite for object-oriented analysis and design). *Object Oriented Systems 3*, 143–158.

HENRY, S. AND KAFURA, K. 1981. Software structure metrics based on information flow. *IEEE Transactions on Software Engineering 7(5)*, 510–518.

JANZEN, D. AND DE VOLDER, K. 2003. Navigating and querying code without getting lost. In *Proceedings 2nd Int. Conf. on Aspect-Oriented Software Development (AOSD)*. ACM Press, Boston, MA, 178–187.

JOHNSON, R. 2003. *J2EE Design and Development*. Wiley Publishing, Indianapolis, IN.

KOSCHKE, R. AND QUANTE, J. 2005. On dynamic feature location. In *Proceedings 20th IEEE/ACM International Conference on Automated Software Engineering (ASE 2005)*. ACM, Boston, MA, 86–95.

KRINKE, J. 2006. Mining control flow graphs for crosscutting concerns. In *AAA Workshop at the 13th Working Conference on Reverse Engineering (WCRE 2006)*. IEEE Computer Society, Washington, DC, USA, 334–342.

LADDAD, R. 2003a. Aspect-oriented refactoring. `www.theserverside.com`.

LADDAD, R. 2003b. *AspectJ in Action - Practical Aspect Oriented Programming*. Manning Publications Co., Greenwich, CT.

LIPPERT, M. AND LOPES, C. 2000. A study on exception detection and handling using aspect-oriented programming. In *Proceedings of the 22nd International Conference on Software Engineering (ICSE)*. ACM Press, Boston, MA, 418–427.

MARIN, M. 2004. Refactoring JHotDraw's Undo concern to AspectJ. In *Proc. of WCRE Workshop on Aspect Reverse Engineering (WARE)*. CWI, Amsterdam, The Netherlands.

MARIN, M., VAN DEURSEN, A., AND MOONEN, L. 2004a. Identifying aspects using fan-in analysis. In *Proceedings of the 11th Working Conference on Reverse Engineering (WCRE2004)*. IEEE Computer Society Press, Los Alamitos, CA, 132–141.

MARIN, M., VAN DEURSEN, A., AND MOONEN, L. 2004b. Identifying aspects using fan-in analysis. Tech. Rep. SEN-R0413, CWI.

MARIN, M., MOONEN, L., AND VAN DEURSEN, A. 2005. A systematic aspect-oriented testing and refactoring process, and its application to JHotDraw. Tech. Rep. SEN-R0507, CWI.

MARIN, M., MOONEN, L., AND VAN DEURSEN, A. 2006. A common framework for aspect mining based on crosscutting concern sorts. In *Proceedings of the 13th Working Conference on Reverse Engineering (WCRE 2006)*. IEEE Computer Society, Washington, DC, USA, 29–38.

MESBAH, A. AND VAN DEURSEN, A. 2005. Crosscutting concerns in J2EE applications. In *Proceedings of the 7th IEEE International Symposium on Web Site Evolution*. IEEE Computer Society, Los Alamitos, CA, 14–21.

MONTEIRO, M. 2004. Catalogue of refactorings for AspectJ. Tech. Rep. UM-DI-GECSD-200401, Universidade do Minho.

MOODIE, M. 2005. *Pro Jakarta Tomcat 5*. Apress, Berkely, CA.

MURALI, T., PAWLAK, R., , AND YOUNESSI, H. 2004. Applying aspect orientation to J2EE business tier patterns. In *Proc. of the 3rd AOSD Workshop on Aspects, Components, and Patterns for Infrastructure Software (ACP4IS)*, Y. Coady and D. Lorenz, Eds. University of Victoria, Victoria, Canada.

MURPHY, G. C., GRISWOLD, W. G., ROBILLARD, M. P., HANNEMANN, J., AND LEONG, W. 2005. Design recommendations for concern elaboration tools. In *Aspect-Oriented Software Development*, R. E. Filman, T. Elrad, S. Clarke, and M. Akşit, Eds. Addison-Wesley, Boston, 507–530.

RICH, C. AND WILLS. L. M. 1990. Recognizing a program's design: A graph-parsing approach. *IEEE Softw. 7*, 1, 82–89.

ROBILLARD, M. AND MURPHY, G. 2002. Concern graphs: Finding and describing concerns using structural program dependencies. In *24th International Conference on Software Engineering (ICSE)*. ACM press, Boston, MA.

SHEPHERD, D., GIBSON, E., AND POLLOCK, L. 2004. Design and evaluation of an automated aspect mining tool. In *Software Engineering Research and Practice*. CSREA Press, Las Vegas, NV, 601–607.

SHEPHERD, D., PALM, J., POLLOCK, L., AND CHU-CARROLL, M. 2005. Timna: A framework for automatically combining aspect mining analyses. In *Proceedings Conference on Automated Software Engineering (ASE'05)*. IEEE Computer Society, Los Alamitos, CA.

SOMMERVILLE, I. 2004. *Software Engineering*, 7th ed. Pearson, NJ.

SUTTON, S. M. AND ROUVELLOU, I. 2005. Concern modeling for aspect-oriented software development. In *Aspect-Oriented Software Development*, R. E. Filman, T. Elrad, S. Clarke, and M. Akşit, Eds. Addison-Wesley, Boston, Chapter 21, 479–505.

TARR, P., HARRISON, W., AND OSSHER, H. 2004. Pervasive Query Support in the Concern Manipulation Environment. Tech. Rep. RC23343 (W0409-135), IBM Research. Sept.

TONELLA, P. AND CECCATO, M. 2004a. Aspect mining through the formal concept analysis of execution traces. In *Proceedings 11th Working Conference on Reverse Engineering (WCRE 2004)*. IEEE Computer Society, Los Alamitos, CA.

TONELLA, P. AND CECCATO, M. 2004b. Migrating interface implementation to aspect-oriented programming. In *Proceedings International Conference on Software Maintenance (ICSM 2004)*. IEEE Computer Society, Los Alamitos, CA, 220–229.

TOURWÉ, T. AND MENS, K. 2004. Mining aspectual views using formal concept analysis. In *Proc. of the Fourth IEEE International Workshop on Source Code Analysis and Manipulation (SCAM 2004)*. IEEE Computer Society, Chicago, Illinois, USA.

VICKERS, P. 2002. Why finalizers should (and can) be avoided. IBM developersWorks articles. `www-136.ibm.com/developerworks/java/`.

WILDE, N. AND SCULLY, M. C. 1995. Software reconnaissance: mapping program features to code. *Journal of Software Maintenance 7*, 1, 49–62.

WILLS, L. M. 1990. Automated program recognition: A feasibility demonstration. *Artificial Intelligence 45*, 1–2 (Sept.), 113–171.

XIE, X., POSHYVANYK, D., AND MARCUS, A. 2006. 3D visualization for concept location in source code. In *28th International Conference on Software Engineering (ICSE 2006)*. ACM, Boston, MA, 839–842.

ZHANG, C. AND JACOBSEN, H.-A. 2003. Quantifying aspects in middleware platforms. In *Proc. 2nd International Conference on Aspect-Oriented Software Development (AOSD)*. ACM Press, Boston, MA, 130–139.

ZHANG, C. AND JACOBSEN, H.-A. 2004. PRISM is Research in aSpect Mining. In *Companion to the 19th Annual ACM SIGPLAN Conference on Object-Oriented Programming, Systems, Languages, and Applications, OOPSLA*. ACM, Boston, MA, 20–21.

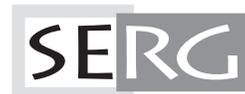